\pdfoutput=1
\documentclass[aps,prb,twocolumn,superscriptaddress]{revtex4}
\usepackage{graphicx}

\begin{document}

\title{Surface defects and conduction in polar oxide heterostructures}

\author{N. C. Bristowe}
\affiliation{Theory of Condensed Matter Group,
             Cavendish Laboratory, University of Cambridge, 
             JJ Thomson Ave, Cambridge CB3 0HE, UK}
\affiliation{Department of Earth Sciences, University of Cambridge, 
             Downing Street, Cambridge CB2 3EQ, UK}
\author{P. B. Littlewood}
\affiliation{Theory of Condensed Matter Group,
             Cavendish Laboratory, University of Cambridge, 
             JJ Thomson Ave, Cambridge CB3 0HE, UK}
\author{Emilio Artacho}
\affiliation{Department of Earth Sciences, University of Cambridge, 
             Downing Street, Cambridge CB2 3EQ, UK}
\affiliation{Donostia International Physics Centre, Universidad del Pais
             Vasco, 20080 San Sebastian, Spain}

\date{\today}

\begin{abstract}
  The polar interface between 
LaAlO$_{3}$ and SrTiO$_{3}$
has shown promise as a field effect transistor,
with reduced (nanoscale) feature sizes and potentially added functionality
over conventional semiconductor systems.
  However, the mobility of the interfacial two-dimensional electron gas
(2DEG) is lower than desirable.
  Therefore to progress, the highly debated origin of the 2DEG must be understood.
  Here we present a case for surface redox reactions as the origin of the 2DEG, in particular
surface O vacancies, using a model supported by first principles
calculations that describes the redox formation.
  In agreement with recent spectroscopic and transport measurements,
we predict a stabilization of such redox processes (and hence Ti 3$d$ occupation)
with film thickness beyond a critical value, which can be 
smaller than the critical thickness for 2D electronic conduction, 
since the surface defects generate trapping potentials that will affect the interface electron mobility. 
  Several other recent experimental results, such as lack of core level broadening and shifts, 
find natural explanation.
  Pristine systems will likely require changed growth conditions or 
modified materials with a higher vacancy free energy.

\end{abstract}


\maketitle

\section{Introduction}

 Complex oxides offer the potential to replace 
conventional semiconductors in a range of devices
due to reduced feature sizes and added functionality
(see e.g. ref.~\cite{Mannhart2010,Takagi2010}).
  The polar interface between 
LaAlO$_{3}$ (LAO) and SrTiO$_{3}$ (STO)
~\cite{Ohtomo2004} 
has shown promise as a field effect transistor
~\cite{Thiel2006,Cen2008}.
  One problem hindering its development is the low mobility
of the interface two-dimensional electron gas (2DEG).
  To progress, the origin of the 2DEG must be understood.
Explanations proposed to date can be classed into three
categories:
  ($i$) electron transfer countering the ``polar catastrophe"
\cite{Nakagawa2006}, 
  ($ii$) doping through O vacancies in the film 
\cite{Cen2008,Li2009,Zhong2010,Zhang2010} or in the substrate 
\cite{Herranz2007,Siemons2007,Kalabukhov2007}, and 
  ($iii$) cation intermixing at the interface
\cite{Nakagawa2006,Willmott2007,Qiao2010}.
  Growth conditions have been shown to affect the observed behavior,
particularly O$_2$ partial pressure and post-annealing treatments,
affecting carrier density and confinement to the interface (as opposed
to conduction through the STO substrate) \cite{Siemons2007,Thiel2006}.

  The polar catastrophe~\cite{Nakagawa2006} arises from a 
polarization discontinuity between the non-polar STO substrate 
and polar LAO film ~\cite{Stengel2009}.
  In LAO films grown on TiO$_2$ terminated STO substrates
this polar discontinuity induces a divergence of the electric 
displacement field ($\vec \nabla \cdot \vec D$)
in the pristine system equal to considering 
the effective polarization response and net effective charges 
 $\sigma_c$ of precisely $+e/2$ and $-e/2$ per formula unit at the interface 
and surface, respectively~\cite{Stengel2009,Bristowe2009,
Bristowe2011}.  
  The electrostatic potential then builds up across the LAO film,
accumulating electrostatic energy.
  To counter it a charge transfer between the interface and 
the surface is required.
  A mechanism intrinsic to the pristine system is given by the 
transfer of electrons from the surface to the interface. 
  From band bending arguments, electrons are transferred 
from the O 2$p$ at the top of the valence 
band at the LAO surface, to the Ti 3$d$ conduction band at the 
interface, once the potential drop across the LAO layer reaches 
the effective band gap, which is calculated to happen for a LAO
thickness of five unit cells (see ref.~\cite{Chen2010} and within).
  However the absence of a 2D hole gas at the surface and 
the observation of populated Ti 3$d$ states for films as thin 
as one or two bilayers~\cite{Sing2009,Berner2010,Fujimori2010} raises 
doubts about this mechanism. 

  Two important points should be kept in mind.
  Firstly, the samples are generally not kept or measured in 
vacuum.
  The surface can thus be covered by adsorbants like water,
and it will likely be far from
an ideal surface termination, possibly including chemical 
alterations such as the hydroxylation seen on many wet oxides.
  The second point is more central, however: whatever the chemistry, 
the relevant electrostatics across the film can only be affected 
by processes in which charge is altered at each side of the film, 
either by charge transport across it, or by charges  arriving
at either side from external reservoirs. 
  If there are no such external sources and 
the chemical processes are confined to the surface, the 
remaining possibility is that of surface redox processes.
  They transform surface bound charge into free-carrier 
charge, the electrons or holes then being free to move to 
the buried interface.
  The clearest and quite relevant example is that of 
oxygen vacancy formation whereby surface O$^{2-}$ anions
transform into O$_2$ molecules, releasing two electrons 
to the $n$ interface, as illustrated in Fig. 1
 \cite{Cen2008,Li2009,Zhong2010,Zhang2010}.
  Surface protonation \cite{Son2010,Sautet2009} is an analogous 
process in which the surface is also reduced by the oxidation of 
O$^{2-}$ into O$_2$, although in this case the process 
depends on the presence of water,
H$_2$O$\rightarrow$ 1/2 O$_2 + 2e^-+2$H$^+$,
and both the energetics and kinetics will be different
from the previous one \cite{Son2010}. 
  The protons attach to the surface O atoms, while
the electrons are again free to go to the $n$ interface
(note that the non-redox hydroxylation,
H$_2$O$\rightarrow$ OH$^- +$H$^+$, does not
affect the electrostatics across the film).

  In section II we present a model for the formation of
surface redox processes. 
  We then apply the model to the LAO/STO system in section III 
using parameters obtained from first-principles calculations (see appendix).
  We show that, for such a 
surface redox process,
  ($i$)  the process is favored by film thickness 
through the effect of the electric field across the film;
  ($ii$) the growth of the density of the related surface defects with
film thickness is predicted, showing a minimum critical thickness;
  ($iii$) 
the model is in excellent agreement with first principles calculations 
and reasonable agreement with what is observed experimentally for Ti 3$d$ occupation;
  ($iv$) the potential drop across the film does not change 
substantially with thickness once vacancies start to appear;
  ($v$) for thin films the carriers at the interface are still trapped 
by the electrostatic potential generated by the vacancies in 
deep double-donor states: a strongly disordered two-dimensional
electron system;
  ($vi$) as thickness increases the levels become shallower
and the number density increases, 
closing the gap to the conduction band;
  ($vii$) the conduction onset is therefore at a higher film thickness
than the one for interfacial carrier population.
  Finally we note that the model can also be applied to 
$p$-type interfaces and ultra thin ferroelectric films, 
where alteration of the surface chemistry has 
also been proposed as a possible screening mechanism (see for 
example ref.~\cite{Wang2009}).
 
\section{Model}

  We consider a pristine polar thin film on a non-polar 
substrate defining a $n$-type interface (see Fig. 1), and a 
surface reduction process (the model can be trivially 
extended to $p$-type interfaces and surface oxidizing
reactions, see below, and to ferroelectric thin films).
  The introduction of a surface defect via a redox reaction 
produces a donor level in the gap. 
  The defect provides $Z$ electrons (two for an O vacancy) 
that can transfer to the interface (see Fig. 1), 
thus contributing to the screening of the polarization or 
compositional charge $\sigma_c$ at either side of the 
polar film ($\pm 0.5e$/f.u. in LAO ~\cite{Stengel2009,Bristowe2009,
Bristowe2011}).
  
  Considering this electron migration, we can model the 
formation energy of one such surface defect, $E_f$, 
in the presence of an area
density $n$ of surface defects as
\begin{equation} 
E_{f}(n) = C +  E_{\cal E}(n) + \alpha n  ,
\end{equation}
where we have separated an electrostatic term associated 
with the internal electric field in the film, $E_{\cal E}(n)$,
from a surface/interface chemistry term, $C$, and a term accounting
for defect-defect interactions other than electrostatic
in a mean-field sense.
  It can be seen as arising from 
\begin{equation} 
E_{f}(n) = E_{f}^{0} - Z(W-E_{CD}) + E_{\cal E}(n) + 
\displaystyle\sum\limits_{i}^{others} J_{i}r_{i} ,
\end{equation}
where $E_{f}^{0}$ is the formation energy of an 
isolated surface defect in the absence of a field across 
the film, and where $W$ and $E_{CD}$ are defined in 
Fig. 1~\footnote{A valence band offset has been omitted 
for clarity since it is known to be small, but can also be 
included alongside the
conduction band offset without affecting the model}.
  $E_{f}^{0}$ (and thus $C$) depends on the particular
surface chemical process, and the
reference chemical potential for the relevant redox 
counterpart species in the environment, e.g. 
$\mu_{\rm O_2}(P,T)$, which depends on experimental
conditions. 

\begin{figure}[t]
\includegraphics[width=0.45\textwidth]{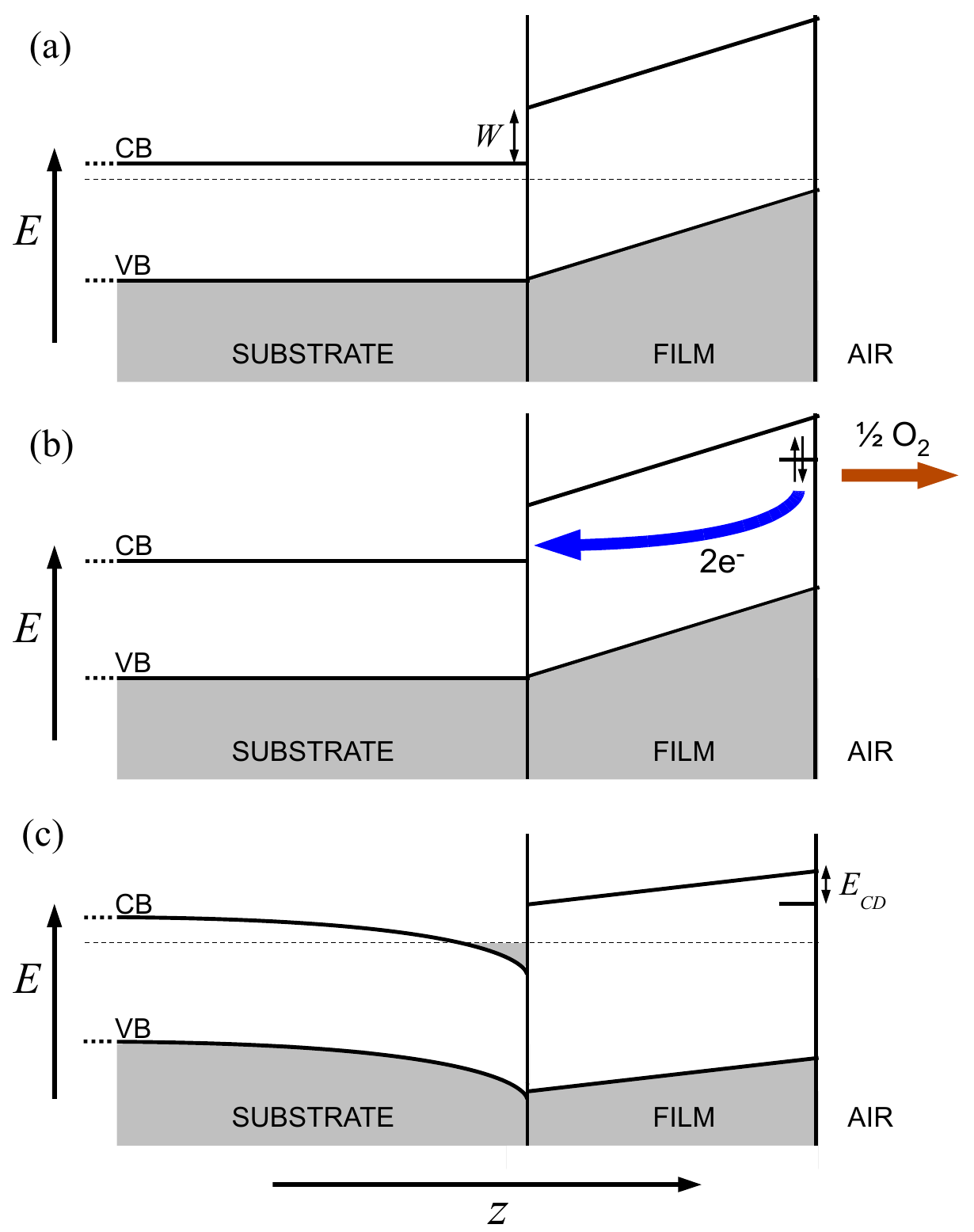}
\caption{\label{BANDS}{(Color online) Schematic band diagram 
of an interface between a polar film and a non-polar substrate
along the normal direction, $z$. 
  (a) The pristine system under the critical film thickness.
  (b) The creation of a donor state at the surface via a redox reaction
  and subsequent electron transfer.
   (c) The reconstruction reduces the film's electric field.}}
\end{figure}

  Taking Eq. 1, the surface excess energy for a given
area density of surface defects is then
\begin{equation} 
\Omega(n) = \int_0^n \! \! \! E_f(n') {\mathrm d} n' = Cn +  \Omega_{\cal E}(n) + \frac{1}{2} \alpha n^2  .
\end{equation}
  The key of the proposed energy decomposition is 
that the $\Omega_{\cal E}$  term is simply the energy gain 
of partly discharging a capacitor, which is
\begin{equation} 
\Omega_{\cal E}(n) = \frac{d}{2\epsilon} \left [ (\sigma_c - \sigma_v)^2 - 
\sigma_c^2 \right ],
\end{equation}
%
where $d$ is the film thickness, $\epsilon$ is the LAO
dielectric constant, $\sigma_c$ is the compositional charge, 
and $\sigma_v= n(Ze)$ is the charge density of the carriers confined to the
interface (note that these electrons 
may not all be mobile, as discussed below).
Eq. 4 assumes no screening by electronic reconstruction, which is right
if the onset for defect stabilisation happens earlier than the one for
electronic reconstruction.
A complete description of all possible regimes will be presented elsewhere
~\cite{Bristowe2010}.
We limit ourselves to the regime given by Eq. 4 since a wider discussion
of the model is irrelevant here.
%
%

  The equilibrium defect density is determined by finding 
the value that minimizes $\Omega$.
  Taking Eqs.~3 and 4, 
\begin{equation} 
n = \frac{d\,Ze\,\sigma_c - C\epsilon}{(Ze)^2d + \alpha \epsilon} .
\end{equation}
  A critical thickness arises for defect stabilization, 
\begin{equation} 
d_c = C \epsilon/(Ze \sigma_c) ,
\end{equation}
$n$ tending to $\sigma_c / Ze$ for large $d$, which is the 
value required to completely screen the film's intrinsic polarization.

\section{Discussion}

\subsection{Interface carrier density}

  We now consider the specific case of O vacancy formation
at the surface as the most prominent candidate redox process
\cite{Cen2008,Li2009,Zhang2010}.
  Fig 2a) shows quantitative agreement between 
the model's $\bar {E}_{f} = {1 \over n} \Omega(n) = {1
\over n} \int_0^n E_f {\mathrm d} n'$
and first principles calculations of 
the surface vacancy formation energies in ref.~\cite{Li2009}
on the full LAO/STO structure.
  $\bar E_f$ is the right magnitude to compare to
  first-principles results since it accounts for the energy difference between
  the system with a given concentration of surface defects ($n$) and the
pristine
  system ($n=0$), per surface defect, thus $\Omega(n)/n$.
The physical constants used in the model were determined 
independently by separate DFT calculations (see appendix)
and then compared to DFT results for films of varying thickness (Fig. 2a)).
  The predicted behavior of $n$ (and $\sigma_v$) in 
LAO/STO is shown in Fig. 2b), where it is compared with Ti  3$d$ 
occupation (both trapped and mobile) as measured with HAXPES~\cite{Sing2009}.  
  Two bands are plotted, the colored one
uses a range in $C$, the striped one a 
range in $\alpha$ and $\epsilon$
(see appendix).
  The bulk dielectric constant of LAO is about 24~\cite{Krupka}.
  It may be substantially different for a strained ultra-thin film (see 
ref.~\cite{Chen2010} and within) and so the range  $21 < \epsilon 
< 46$ has been considered. 

\begin{figure}[t]
\includegraphics[width=0.4\textwidth]{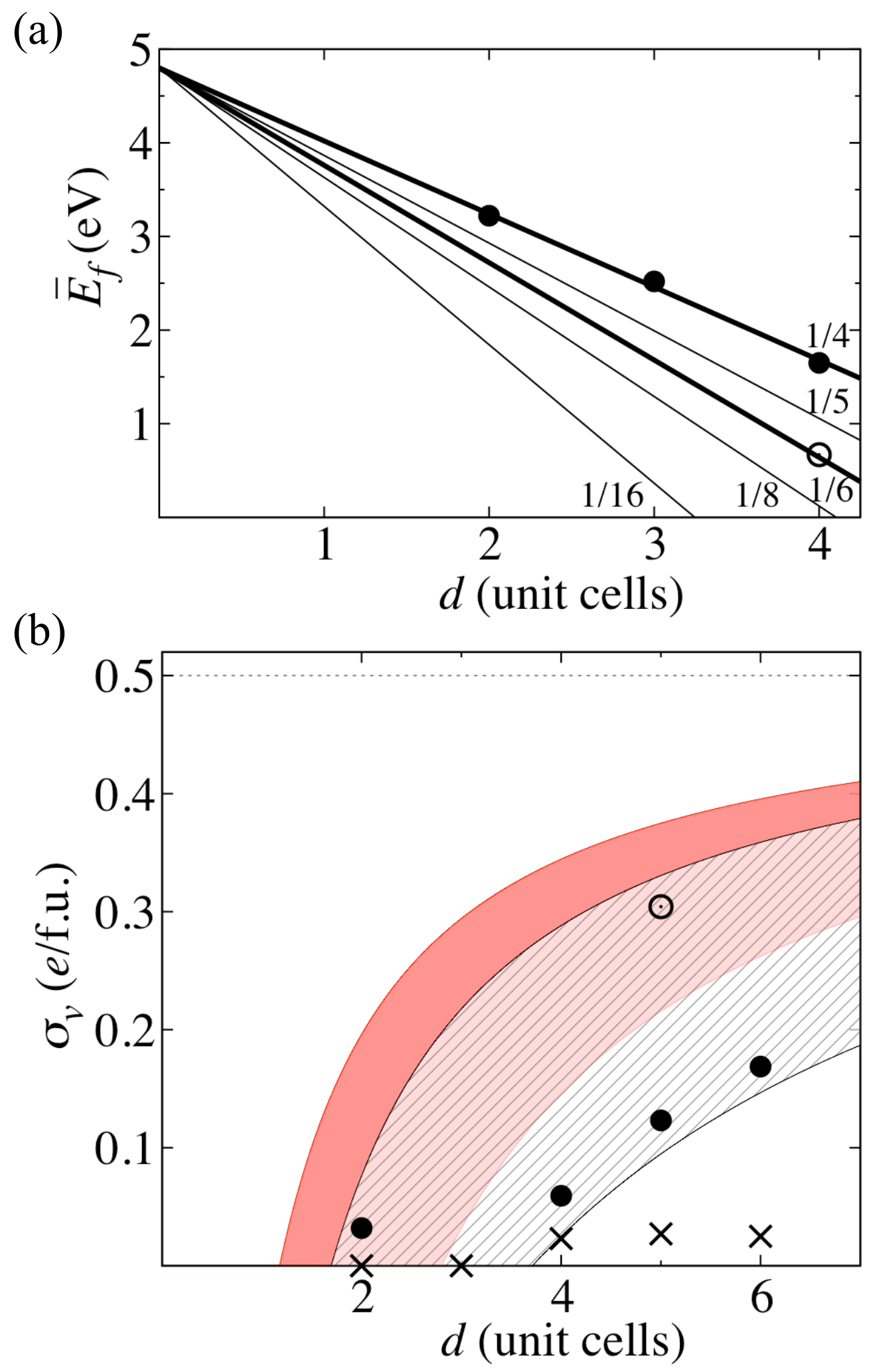}
\caption{\label{BAND}{(Color online) 
a) 
Defect formation energy, $\bar {E}_{f}$
(see definition in text)
versus LAO film thickness $d$
for various vacancy densities $n$.
The model (lines) is compared with
 the DFT calculations (circles) of ref.~\cite{Li2009} 
 of the surface vacancy formation energy on the 
 full LAO/STO structure
 (see appendix for the determination of the
 model parameters) 
b)
 Equilibrium area density of interface
carriers $\sigma_v$ versus $d$. 
  The red (grey) band is the model
prediction for 2.1 eV $< C < 5.0$ eV, $\epsilon=25$ and 
$\alpha=0.8$ eV/(vac/f.u.)$^2$. 
  The striped band is for $C=3.6$ eV, 
$21 < \epsilon < 46$ and $0 <\alpha<8$ eV/(vac/f.u.)$^2$.
  The circles indicate the Ti 3$d$ occupation as measured 
with HAXPES in ref.~\cite{Sing2009}.
  Open circle indicates the sample was not annealed.
  The crosses indicate the carrier density from Hall measurements in 
  ref.~\cite{Thiel2006}.}}
\end{figure}
 
  The agreement in Fig. 2b) between model and
experiment is only qualitative given the 
ambiguities in some of the magnitudes of key parameters 
defining the problem, most notably the chemical potential 
of O$_2$ in experimental conditions.
  Despite this, the model 
predicts a critical thickness for the appearance 
of carriers at the interface for a LAO film thickness 
below the five unit cells predicted by the purely electronic
mechanism.
  Other qualitative features observed but not understood
in this system also find a natural explanation (below).

\subsection{Electric field in LAO: pinning of potential drop}

 The electrostatic potential drop across the LAO film is
\begin{equation} 
V = (\sigma_c - \sigma_v)d / \epsilon  .
\end{equation}
  Substituting $\sigma_v$, the drop is essentially 
independent of thickness, $V\approx C/(Ze)$, when 
the vacancy-vacancy interaction is small, 
$\alpha \ll (Ze)^2 d / \epsilon$.
  Using the parameters for LAO/STO 
the difference in potential drop per LAO layer added is 
between 0.0 and 0.2 eV/f.u., much smaller than the predicted 
thickness dependence for the electronic screening model,
and consistent with recent XPS measurements, which show 
no core level broadening with film thickness~\cite{Segal2009}. 
The pinning of $V$ is also consistent with reduced cation-anion
relative displacement with increasing LAO thickness as measured by SXRD~\cite{Pauli2011}.

\subsection{Onset of conduction: electron trapping}

  The redox processes proposed above explain the absence
  of hole-mediated transport at the surface, while 
electrons allow 2D conduction at the interface.
  An important  observation that remains unexplained,
however, is the fact that the onset of interfacial Ti 3$d$ 
occupation, as measured with HAXPES~\cite{Sing2009,Berner2010,Fujimori2010}, 
happens at lower film thickness than the onset for interfacial
2D conduction~\cite{Thiel2006}.
  It has been suggested that the 2DEG lies in several Ti 3$d$ 
sub-bands, some of which are not mobile due to Anderson 
localization~\cite{Popovic2008}. 
  Whether Anderson localization occurs on an energy scale 
as high as room temperature depends on the energy scale 
of the disorder distribution. 
  The surface defects associated to the redox processes 
represent point sources of effective charge, very much as a 
dopant in a semiconductor \cite{Bristowe2011}, e.g. $+2$ 
$e$ for an O vacancy.
  They then generate trapping potentials for the carriers
at the interface plane of the form
$ 
V_{\rm trap} = Z e^2 / \epsilon \sqrt{\rho^2+d^2} ,
$ 
in atomic units, where $d$ is the film thickness and 
$\rho^2 = x^2+y^2$ corresponds to the radial variable in 
the plane.
  This potential is sketched in Fig. 3 for several $d$ values.
  Its depth decays with thickness as $1/d$.
    Fig. 3 shows estimates of the ground state electron level
associated to the double donor state arising at the interface
due to an O vacancy at the surface.
  These trapped interface levels may be the 'in-gap states' seen in a 
recent spectroscopic study \cite{Drera2011}.
  For a thin film the traps are deep and few, but as it grows thicker, 
the donor states become shallower {\it and} the area 
density of traps grows, as illustrated in the inset of Fig. 3.
  A transition from insulating to conducting behavior is
thus expected at a larger film thickness than the critical
thickness for surface defect stabilization.
  With growing thickness, not only dopant levels tend to 
overlap as in a degenerate semiconductor, but the 
doping-level band is pushed towards the conduction 
band.
  For $Z=2$, as for O vacancies, the physics of this
transition is that of band overlap and disorder, since all
dopant states are doubly occupied.
  If the mechanism involves $Z=1$ defects, as in the hydroxylation
case, the transition will be rather Mott-Anderson, as each
dopant state is singly occupied.
  The different phenomenologies could be used to 
ascertain on the mechanism.
  The surface potential distribution from charged defects is consistent
  with a recent Kelvin probe force microscopy study~\cite{Popok2010}. 
 
\begin{figure}[t]
\includegraphics[width=0.45\textwidth]{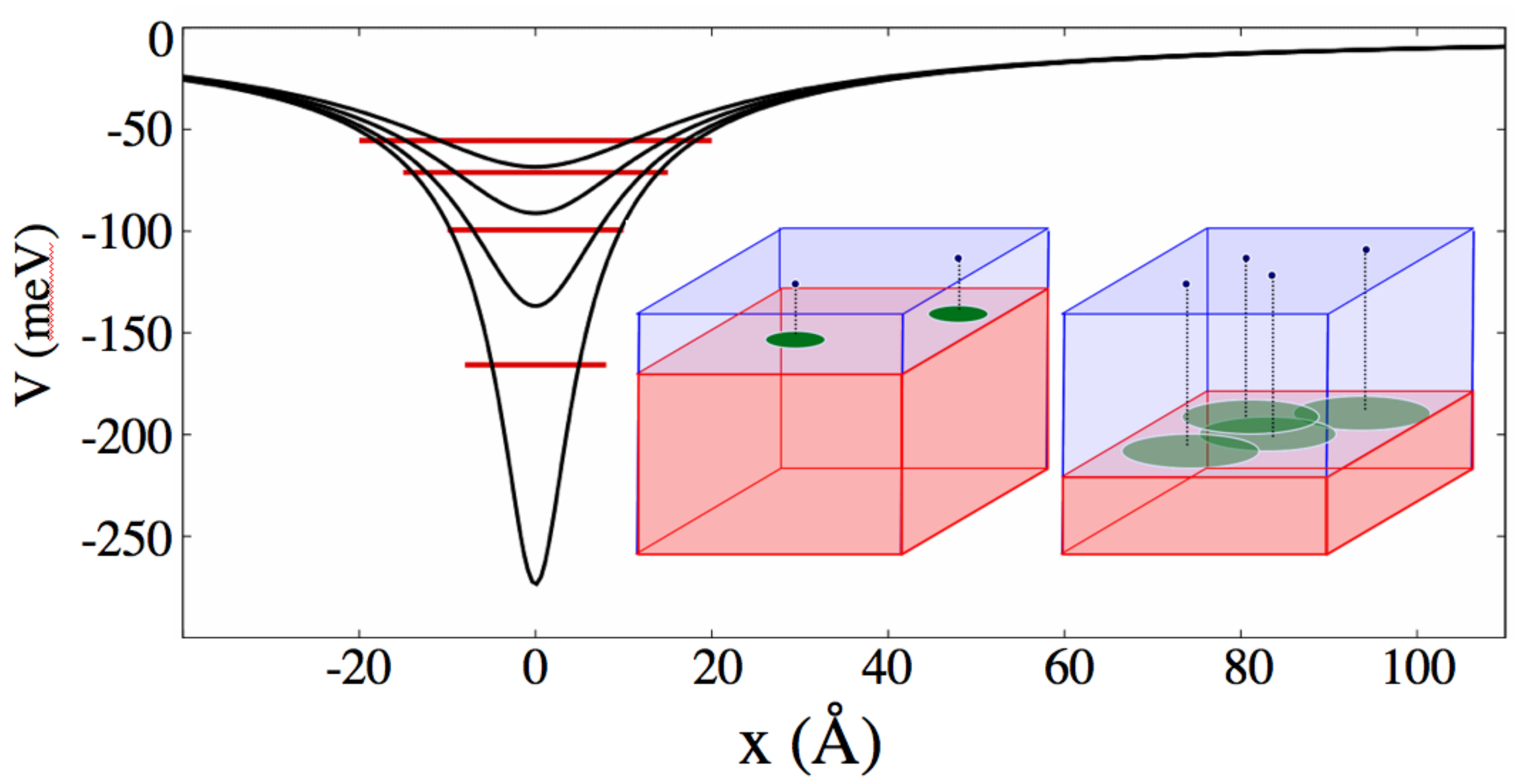}
\caption{\label{TRAP}{(Color online) Trapping potential $V$ created by a surface O vacancy 
as seen by interface electrons
versus distance within the interface plane $x$ ($x$=0 is directly below the vacancy)
for film thickness $d=a$ (deepest)$, 2a, 3a,4a$ (shallowest), with
 harmonic estimates of corresponding donor ground states
(taking $m_{\rm eff} = 3 m_e$ \cite{Dubroka2010}). Inset: Sketch 
of range and density of trapped states.}}
\end{figure} 
 
\section{Final remarks and conclusions}
 
  A recent AFM study of LAO/STO has proposed the mechanism for 
conductivity switching~\cite{Thiel2006,Cen2008} as the writing 
of surface charge~\cite{Xie}.
  Applying a biased tip to the surface alters the field across the LAO 
film which either increases or decreases the stability of vacancies 
(and hence $\sigma_v$) depending on the sign of the bias.
  An implication of this observation is that the kinetics
for these redox processes is accessible at room $T$ as
used in these experiments, not only the much higher $T$ used for
growth.
 
  The model proposed can also be used for $p$ interfaces, holes,
acceptor levels and surface oxidation processes.
  This would be the case for LAO grown on SrO terminated 
STO.
  It is less symmetric than it seems, however, since, in addition
to different energetics and chemical potentials, the large
conduction band offset at the interface ($W$ in Fig. 1) favors
the situation for electrons towards $n$ interfaces much more 
than the much smaller valence band offset for holes and 
$p$-interfaces.
  For thin film ferroelectrics with outwards (inwards) polarization on metallic 
substrates, the important alignment becomes the acceptor (donor) level 
with the metal fermi level.
  This could be behind the stability of switchable ultra 
thin ferroelectric films under open circuit conditions
(see e.g. ref.~\cite{Garcia2009,Wang2009}).


  We conclude that in LAO/STO, the onset of electrostatic modulation 
doping is precluded by the thermodynamic creation of surface 
defects and thus carrier mobilities produced by this method will be 
much lower than at a pristine interface. 
  Intrinsic systems will likely require changed growth conditions or 
modified materials with a higher vacancy free energy.



\begin{acknowledgments}
  We acknowledge H Hwang, M Pruneda and M Stengel for valuable discussions,
the support of EPSRC and the computing 
resources of CamGRID at Cambridge University and the 
Spanish Supercomputer Network (RES).
\end{acknowledgments}


\appendix*
\section{Model parameters}

Here we describe the determination of the parameters,  $\alpha$, $\epsilon$ and $C$ used in the model in Fig. 2 of the main paper.
The parameters used in the model were independently determined
from first principles in appropriate LAO-based systems (see below), and then the model
was checked against DFT calculations of vacancy formation
energies in the full LAO/STO system as a function of thickness (Fig. 2a)).
When comparing the model with experiment (Fig. 2b)), we account for inaccuracies of DFT and the ambiguity of the experimental
chemical potential in the determination of each of these parameters.

\subsection{The vacancy-vacancy interaction term, $\alpha$}

The defect-defect term was defined in Eq. 1 to include interactions other than electrostatic. 
Therefore to determine $\alpha$ we performed first principles calculations (see ref~\cite{Bristowe2009} for the method) 
of the charge neutral defect, i.e. oxygen vacancies in 'bulk' LAO which include the double donor electrons.
One oxygen vacancy was placed in a simulation cell of 1$\times$1$\times$8, 2$\times$2$\times$8 and 3$\times$3$\times$8 unit cells of LAO to approximate 2D arrays of vacancies of various area densities. 
From the difference in formation energy per vacancy between the three calculations, $\alpha$ was found to be $0.8$ eV/(vac/f.u.)$^2$, which was used in Fig 2a).
$\alpha$ is formally defined as the interaction between vacancies at the film surface, however we believe this bulk value to be a good estimate.
For the comparison with experiment, to account for any error associated with this determination we choose the range  $0 <\alpha<8$ eV/(vac/f.u.)$^2$ in Fig. 2b) of the paper.

\subsection{The dielectric constant,  $\epsilon$}

The dielectric constant,  $\epsilon$, consists of lattice and electronic contributions.
For LAO, we take $\epsilon$=28 as the total for Fig 2a), as consistent with ref.~\cite{Delugas2005}.
When comparing with experiment, we note the error and inconstancy of DFT calculations of $\epsilon$, and additionally the effect of strain as highlighted in ref.~\cite{Chen2010}.
Due to this we choose the range $21 <\epsilon < 46$ for Fig 2b).

\subsection{The surface/interface chemistry term,  $C$}

$C$ consists of three terms,

\begin{equation}
C=E_{f,\mu=0}^{0} +\mu + 2(E_{CV}-W)
\end{equation}

From the electronic structure presented by Li $et$ $al.$, the last term is found to be approx. -1.2 eV.
From first principles calculations of ref.~\cite{Jin-Long2008} the formation energy of an isolated
oxygen vacancy at the surface of LAO (in the absence of a field)
with reference to oxygen in an isolated molecule (1/2$E[O_2]$),
$E_{f,\mu=0}^{0}$, is approx. 6.0 eV.
We define the zero of chemical potential relative to this reference state,
which is appropriate for the DFT comparison
and hence the value of $C$ used in Fig. 2a) is taken as 4.8 eV.

The DFT underestimation of the band gap requires corrections to both $E_{f,\mu=0}^{0}$ (see ref.~\cite{Persson2005}) and $W$ for the comparison with experiment in Fig. 2b).
At this point we note the difficulties and variation in first principles determination of formation energies of donor/acceptor states (see for example ref.~\cite{De'ak2008}).

From ref.~\cite{Persson2005}, the formation energy correction of a donor defect, required due to DFT band gap underestimation is simply:

\begin{equation}
\Delta E_{f}^{0}=Z\Delta E_c
\end{equation}

where $\Delta E_c$ is the change in conduction band edge between LDA and experiment (or corrected DFT).
By comparing the electronic structure presented in Li $et$ $al.$ and ref.~\cite{Xiong2008}, this correction could be as large as 1.0 eV. 
Therefore we take $6.0~\mathrm{eV} < E_{f,\mu=0}^{0} < 7.0~\mathrm{eV}$.


From experimental band alignment~\cite{Yoshimatsu2008} and theoretical calculations determining the gap states~\cite{Xiong2008}, the third term in Eq. (A.1) is approx. 2.0 eV (not 1.2 eV).
Correcting for these DFT errors we take, 

\begin{equation}
4.0~\mathrm{eV}+\mu < C < 5.0~\mathrm{eV}+\mu
\end{equation}

The chemical potential of oxygen in the growth conditions used in ref.~\cite{Sing2009} ($T=1073~$K and $p=2.0\times10^{-8}$ atm) relative to the zero reference defined above
is calculated to be -1.9 eV assuming the environment acts as an ideal gas-like reservoir. 
The effect of post-annealing and cooling to room temperature and pressure is to shift the chemical potential towards zero.
With these limits on the chemical potential and the inequality in Eq. (A.3), the range of $C$ becomes $2.1~\mathrm{eV} < C < 5.0~\mathrm{eV}$, as used in Fig. 2b).


%

\end{document}